\documentclass[preprint,showkeys,secnumarabic,amsfonts,amsmath,amssymb]{revtex4}
\usepackage[dvips]{color}
\usepackage{array}
\usepackage{epsfig}
\usepackage{amsmath}
\usepackage{graphicx}
\begin{document}
\title{Interacting Agegraphic Dark Energy Model in DGP Braneworld Cosmology: Dynamical System Approach}

\author{A. Ravanpak}
\email{a.ravanpak@vru.ac.ir}
\affiliation{Department of Physics, Vali-e-Asr University, Rafsanjan, Iran}
\author{G. F. Fadakar}
\email{g.farpour@vru.ac.ir}
\affiliation{Department of Physics, Vali-e-Asr University, Rafsanjan, Iran}

\date{\today}
\begin{abstract}
A proposal to study the effect of interaction in an agegraphic dark energy model in DGP brane-world cosmology is presented in this manuscript. After explaining the details, we proceed to apply the dynamical system approach to the model to analyze its stability. We first, constrain model parameters with a variety of independent observational data such as cosmic microwave background anisotropies, baryon acoustic oscillation peaks and observational Hubble data. Then, we obtain the critical points related to different cosmological epochs. In particular, we conclude that in the presence of interaction, dark energy dominated era could be a stable point if model parameters $n$ and $\beta$, obey a given constraint. Also, big rip singularity is avoidable in this model.
\end{abstract}

\maketitle

\section{Introduction}

Large-scale observations indicate a late-time cosmic acceleration \cite{Riess}-\cite{Eisenstein} . Pressureless matter which offers standard gravitational attraction cannot speed up the universe. Hence, several scenarios proposed to explain this cosmic accelerated expansion \cite{Nojiri}-\cite{Cai} . The basic idea is to insert a new component with an effective negative pressure into Einstein's energy-momentum tensor to push the universe apart, called dark energy (DE). Various DE candidates have been proposed in the literature, for instance, the cosmological constant \cite{Sahni} and scalar field DE models \cite{Caldwell}-\cite{Caldwell2} .

Understanding the nature of DE in a fundamental theory such as quantum gravity may be more feasible. Holographic DE (HDE) \cite{Horava}-\cite{Wei} and agegraphic DE (ADE) \cite{Cai2} are some other types of DE models that originate from string theory. HDE and ADE are consistent with quantum principle, in the sense that they obey a Heisenberg type
uncertainty relation. Also, both of them predict a time-varying DE equation of state (EoS) and are very successful in explaining observational data. Choosing the event horizon of the universe as the length scale, the late-time acceleration is obtained naturally in HDE model. But, there is an obvious shortcoming in holographic approach because the causality appears in this theory. Since the event horizon is a general concept of space-time and is determined by future evolution of the universe, it only
exists for an evermore accelerated expanding universe \cite{Guberina}-\cite{Wei3} . Therefore, in a similar approach, the author in \cite{Cai2} proposed a new quantum-based DE model as a solution to the causality problem in the holographic scenario, called ADE in which the age of the universe is considered to be the length scale.

In addition to the uncertainty relation of quantum mechanics, ADE models assume that the observed DE, originates from space-time and matter field fluctuations in the universe. In fact, combining K\'{a}rolyh\'{a}zy relation \cite{Karolyhazy} , $\delta t=\zeta t_p^{2/3}t^{1/3}$, in which $\zeta$, is a dimensionless constant of order unity, with the time-energy uncertainty principle, an expression for the energy density of quantum fluctuations in Minkowski space-time could be obtained as \cite{Maziashvili}
\begin{eqnarray}\label{rho}
\rho&\sim&\frac{1}{t_p^2t^2}\sim\frac{M_p^2}{t^2}\label{rho},
\end{eqnarray}
in which $t$, is the length scale and $t_p$ and $M_p$, are the reduced Planck time and Planck mass, respectively.

Although non-interacting ADE model attracted a plenty amount of attention since its birth \cite{Wei4}-\cite{Farajollahi} , investigating the interaction between dark matter and ADE and its consequences was of particular interest for many authors \cite{Wei2}-\cite{Farajollahi2} . In fact, one of the most important motivations for considering the interaction between the dark sectors of the universe is to solve the coincidence problem \cite{Baglaa}-\cite{Wang} .

Independent of what mentioned heretofore, extra dimensional theories in which our four-dimensional (4D) universe is considered as a brane embedded in a higher
dimensional space-time dubbed bulk, have attracted a great deal of attention since nearly the beginning of the current century \cite{Arkani}-\cite{Randall2} . In brane-world scenarios the standard model of particle physics is confined to a 4D brane and only gravity can leak into the bulk. Among various brane-world scenarios the one proposed by Dvali, Gabadadze and Porrati (DGP) which includes an infinite Minkowski bulk is remarkably suitable to explain the late-time acceleration of the universe \cite{Dvali} . With attention to the two different states that the brane can be embedded in the bulk, the DGP model contains two separate branches with distinct characteristics denoted by $\epsilon=\pm1$. The self-accelerating branch with $\epsilon=+1$ that naturally leads to the late-time acceleration and the case with $\epsilon=-1$, that needs a DE component to explain an accelerating phase, called normal branch.

Recently, many authors have studied various DE candidates in a normal DGP model, such as cosmological constant ($\Lambda$DGP)\cite{Sahni2}-\cite{Lazkoz} , scalar fields (SDGP) \cite{Chimento}-\cite{Quiros} , Chaplygin gas \cite{Lopez}-\cite{Rudra} , HDE \cite{Wu}-\cite{Dutta} and ADE \cite{Farajollahi4} . Among them, the authors in \cite{Rudra} and \cite{Dutta} , have considered an interacting model and investigated the effect of interaction in their works. For instance, the authors in \cite{Rudra} , have indicated a greater tendency of the flow to go towards a specific attractor point with gradually increasing value of interaction. Also, in \cite{Dutta} , the authors have shown that the first and the generalized second law of thermodynamics are affected by interaction if one consider the event horizon as the boundary of the universe.

Herein, we would like to investigate the evolution of the vacuum energy on the brane in a normal DGP model according to the agegraphic principle in a general but powerful mechanism called dynamical system analysis. This is an approach that has greatly pervaded the cosmological researches, because it determines the fixed points of the system under consideration which may represent important cosmological solutions \cite{Stachowski}-\cite{Xiao} . We show that if an interacting ADE model is considered, DE dominated era will be a stable point while in a non-interacting case, it is a saddle point \cite{Arvin} .

The paper is organized as follows: In Sec.II, we explain the interacting ADE model in a DGP brane-world cosmology in detail. Sec.III, is dedicated to the dynamical system approach, though it also includes a numerical best-fitting procedure. Sec.IV, includes a discussion about the big rip singularity in our model. Summary and remarks have been expressed in Sec.V.

\section{The ADE in DGP model}

As we mentioned in the introduction, in ADE scenario the age of the universe
\begin{equation}\label{T}
T=\int_0^a\frac{da}{Ha},
\end{equation}
is considered as the length scale in which $a$ and $H$, are the scale factor and the Hubble parameter, respectively. Substituting $T$ for $t$, in Eq.(\ref{rho}), agegraphic energy density can be written as \cite{Cai2}
\begin{eqnarray}\label{rhoD}
\rho_{DE}&=&\frac{3n^2M_p^2}{T^2}.
\end{eqnarray}
Here, $3n^2$ is a numerical factor that parameterizes some uncertainties, such as the effect of curved space-time and the species of quantum fields in the universe. Note that the constant parameter $n$, in ADE model has the same role as $c$, in HDE model. The Friedmann equation on the brane in a normal DGP brane-world model is \cite{Deffayet}
\begin{eqnarray}\label{fried}
H^2+\frac{H}{r_c}&=&\frac{\rho_m+\rho_{DE}}{3M_p^2},
\end{eqnarray}
in which $r_c$, is called crossover length scale which separates 4D and 5D regimes of the model and $\rho_m$, is the dark matter density of the universe on the brane.

In the following we proceed to study the dynamical system analysis of our model. We start by introducing the fractional energy densities as
\begin{equation}
\Omega_m=\frac{\rho_m}{3M_p^2H^2},\quad \Omega_{DE}=\frac{\rho_{DE}}{3M_p^2H^2},\quad \Omega_{r_c}=\frac{1}{4r_c^2H_0^2},
\end{equation}
(Hereafter, the subscript 0, refers to the current value of cosmological parameters). Using the above definitions we can rewrite the Friedmann equation as
\begin{equation}\label{cons}
\Omega_{m}+\Omega_{DE}+\Omega_{DGP}=1,
\end{equation}
in which we have introduced a new dimensionless cosmological parameter related to the extra dimension as $\Omega_{DGP}=2\epsilon\frac{H_0}{H}\sqrt{\Omega_{r_c}}$. Combining Eq.(\ref{rhoD}) with the definition of $\Omega_{DE}$, we can obtain another useful relation as
\begin{eqnarray}\label{omega}
\Omega_{DE}&=&\frac{n^2}{H^2T^2}.
\end{eqnarray}
Also, differentiating Eq.(\ref{rho}) with respect to the cosmic time and using the prior relation we reach to
\begin{eqnarray}\label{rhot}
\dot\rho_{DE} = -2H\rho_{DE}\frac{\sqrt{\Omega_{DE}}}{n}.
\end{eqnarray}
Considering the interaction between the dark sectors of the universe the energy densities no longer evolve independently, rather they satisfy the following equations
\begin{eqnarray}\label{ct1}
\dot{\rho}_{m}&+&3H\rho_{m}=Q,
\end{eqnarray}
\begin{equation}\label{ct}
\dot{\rho}_{DE} + 3H(1+w_{DE})\rho_{DE}=-Q.
\end{equation}
in which $Q$, represents the interaction term which gives the rate of energy exchange between the dark components of the universe. Combining Eq.(\ref{rhot}) with Eq.(\ref{ct}), we find the EoS parameter of the ADE as
\begin{eqnarray}\label{omega2}
w_{DE}&=&-1+\frac{2}{3n}\sqrt{\Omega_{DE}}-\frac{Q}{3H\rho_{DE}}\,.
\end{eqnarray}

There is not a unique form for $Q$, in the literature. It has been considered as a term proportional to $\rho_m$, $\rho_{DE}$, $\rho_m+\rho_{DE}$, $\sqrt{\rho_{DE}\rho_m}$ or $\frac{\rho_{DE}\rho_m}{\rho_{DE}+\rho_m}$ \cite{Li2}-\cite{Feng} . Also, in a model with conformal coupling between a scalar field DE and the matter content, chameleon cosmology, the interaction term naturally appears as a product of $\rho_m$ and the time variation of the scalar field coupling function \cite{Farajollahi2} ,\cite{Farajollahi8} . In the following, we consider the form of interaction as $Q=3\beta H\frac{\rho_{DE}\rho_m}{\rho_{DE}+\rho_m}$, that has been preferred in \cite{Feng} . Therein, by employing the information criteria method the authors have shown that in an interacting HDE model this type of interaction is most favored by Planck 2015 results. $\beta$ is a dimensionless coupling parameter that determines the strength of the interaction. Also, according to our convention $\beta$, has a positive value which means that the ADE decays to dark matter.

\section{Dynamical system approach and stability analysis}

Dynamical system approach is a mathematical tool which gives insights on the long-term behavior of the model under consideration and its evolution near the fixed points, using stability analysis. Generally, a dynamical system of order $n$, is defined as follows:

1. The state of the system at any time $t$, can be represented by $n$ real variables that can be considered as coordinates of a vector in an $n$D space, called phase-space.

2. The time evolution of the system, is represented by a set of first order equations, called equations of motion.

If time does not appear in the equations explicitly, we have a time-independent or an autonomous system which is the case of interest in this approach. In order to perform the stability analysis, one has to introduce some auxiliary variables so that the cosmological equations of motion turn into a self-autonomous dynamical system. To this aim, we introduce the following new dimensionless variables:
\begin{eqnarray}\label{nv}
 x=\sqrt{\frac{\rho_m}{3M_p^2(H^2+\frac{H}{r_c})}}\,,\quad y=\sqrt{\frac{\rho_{DE}}{3M_p^2(H^2+\frac{H}{r_c})}}\,.
\end{eqnarray}
So, Eq.(\ref{fried}), yields the Friedmann constraint as
\begin{equation}\label{constraint}
x^2+y^2=1.
\end{equation}
Because our phase-space variables $x$ and $y$, should be non-negative, and with attention to Eq.(\ref{constraint}), one can obtain the following constraints on phase-space variables: $0 \leq x\leq1$, and $0 \leq y\leq1$. Now, we can rewrite the EoS parameter of ADE, Eq.(\ref{omega2}), in terms of new variables as
\begin{equation}\label{eos2}
    w_{DE}=-1+\frac{2}{3n}yz-\beta x^2\,.
\end{equation}
Also, differentiating Friedmann equation with respect to time $t$, and using Eqs.(\ref{ct1}) and (\ref{ct}), after some calculations we obtain
\begin{equation}\label{ray}
\frac{\dot H}{H^2}=\frac{-3z^2(x^2+\frac{2}{3n}y^3z-\beta x^2y^2)}{z^2+1}\,,
\end{equation}
in which $z=\sqrt{1+\frac{1}{Hr_c}}$. Since both $H$ and $r_c$, are positive, so $z\geq1$. The 4D situation corresponds to the limit $r_c\rightarrow \infty$. Considering the phase-space variables, Eq.(\ref{nv}), using Eqs.(\ref{constraint}) and (\ref{ray}), we obtain the following system of ordinary differential equations in $xy$-plane as
\begin{eqnarray}
  x' &=& -\frac32x+\frac{3}{2}\beta xy^2+\frac32x\left(x^2+\frac{2}{3n}y^3z-\beta x^2y^2\right)\,, \\
  y' &=& -\frac{y^2z}{n}+\frac32y\left(x^2+\frac{2}{3n}y^3z-\beta x^2y^2\right)\,.
\end{eqnarray}
Here, prime means derivative with respect to $\ln a$. Regarding the definitions mentioned above, this system is autonomous, though we have considered the evolution of the system with respect to $\ln a$, instead of $t$.

Now, using stability analysis we want to study the behavior of our system in the vicinity of its critical points. To this aim, we impose the conditions $x'=0$ and $y'=0$ simultaneously, and find the fixed points of our model. But before that, we try to fit our model parameters $n$, $\Omega_{DE0}$, $\Omega_{r_c}$, $H_0$ and the interaction coefficient $\beta$, with observational data. We use $\chi^2$ method for a combination of cosmic microwave background (CMB) data, baryon acoustic oscillations (BAO) data and observational Hubble data (OHD). The results have been shown in TABLE \ref{table:1}.\\

\begin{table}[h]
\caption{Best-fitted values of model parameters} 
\centering 
\begin{tabular}{c|c|c|c|c|c} 
\hline\hline 
model parameters \ & \ $H_0$ \ & \ $\Omega_{DE0}$ \ & \ $n$ \ & \ $\beta$ \ & \ $\Omega_{r_c}$ \\ [1ex]
\hline
best-fitted values & 67 & 0.82 & 14 & 0.18 & 0.0099 \\ 
\hline
\end{tabular}
\label{table:1} 
\end{table}\

TABLE \ref{table:2}, indicates the allowed fixed points which satisfy the constraints on $x$ and $y$. To discuss these critical points we use relations $xz=\sqrt{\Omega_m}$ and $yz=\sqrt{\Omega_{DE}}$, and advert to Eq.(\ref{cons}). At point $A$, $\Omega_{DE}=0$ and $\sqrt{\Omega_m}=z$. We know that $z\geq1$ and the maximum value of $\Omega_m$ is 1. So we conclude that $\Omega_m=1$ and therefore point $A$ demonstrates a matter dominated era. In the same way, point $B$, relates to a DE dominated era. Because at this point we have $\Omega_m=0$ and $\sqrt{\Omega_{DE}}=z$. With attention to the constraint on $z$ and Eq.(\ref{cons}), we find that $\Omega_{DE}$ is at least equal to 1 (because $\Omega_{DGP}$ is always negative). Therefore this point indicates a DE dominated epoch.

\begin{table}[h]
  \caption{Fixed points of the model}
      \begin{tabular}{ccccc}
      \hline\hline
      points & $(x,y)$ & eigenvalues & description & stability \\ [1ex] \hline
      A & (1 , 0) & $\left(\frac{3}{2} , \frac{3}{2}(1+\beta)\right)$ & matter dominated & unstable \\ [1ex] \hline
      B & (0 , 1) & $\left(\frac{2-3n}{n} , \frac{3\beta n-3n+2}{2n}\right)$ & DE dominated & \begin{tabular}{cc} stable  &  if $\beta<1-\frac{2}{3n}$ \\ [0.1ex] saddle & if $\beta>1-\frac{2}{3n}$ \end{tabular}\\ \hline
      \end{tabular}
\label{table:2} 
\end{table}

\begin{figure}
\centering
\includegraphics[width=6cm]{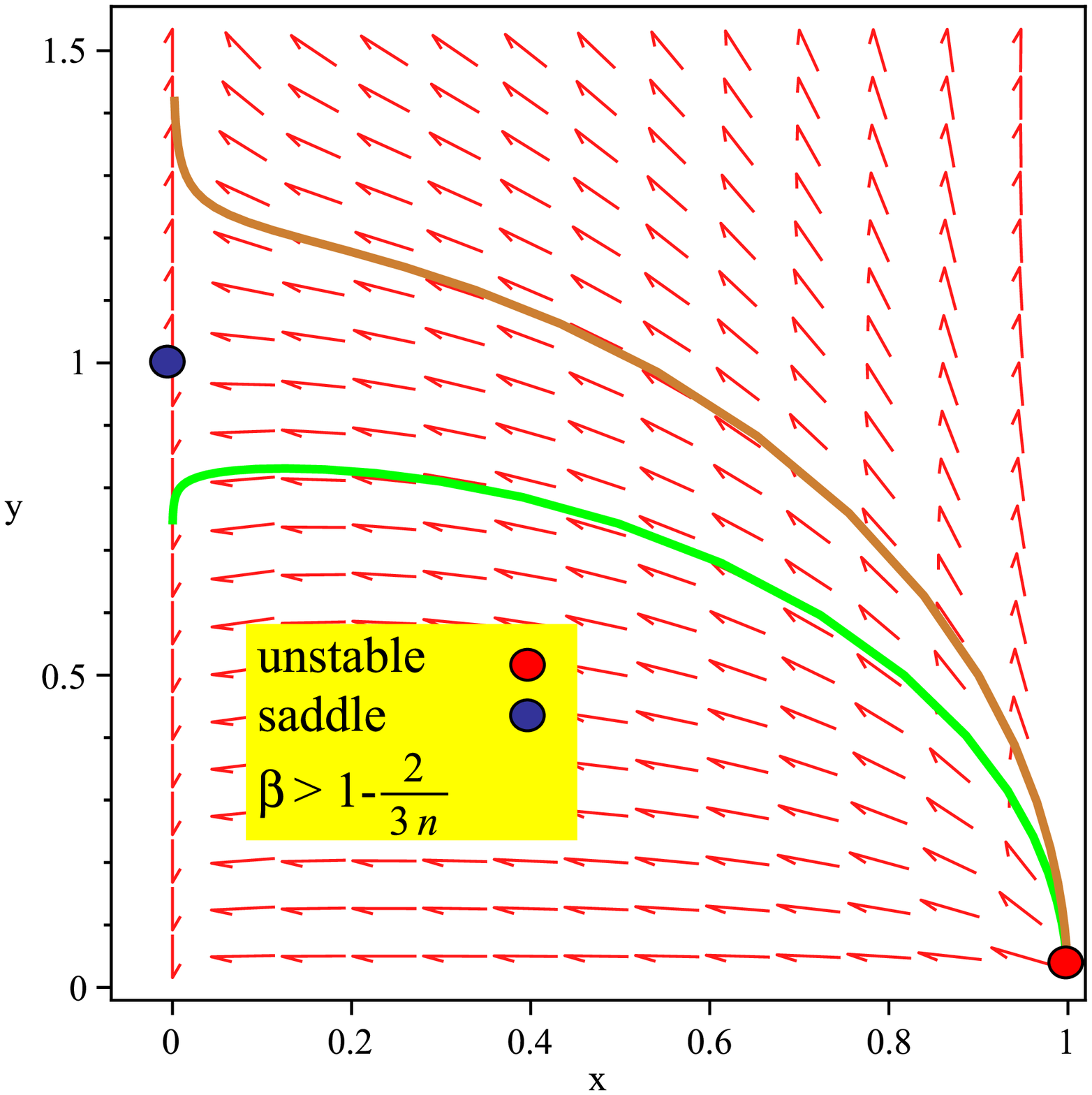}
\includegraphics[width=6cm]{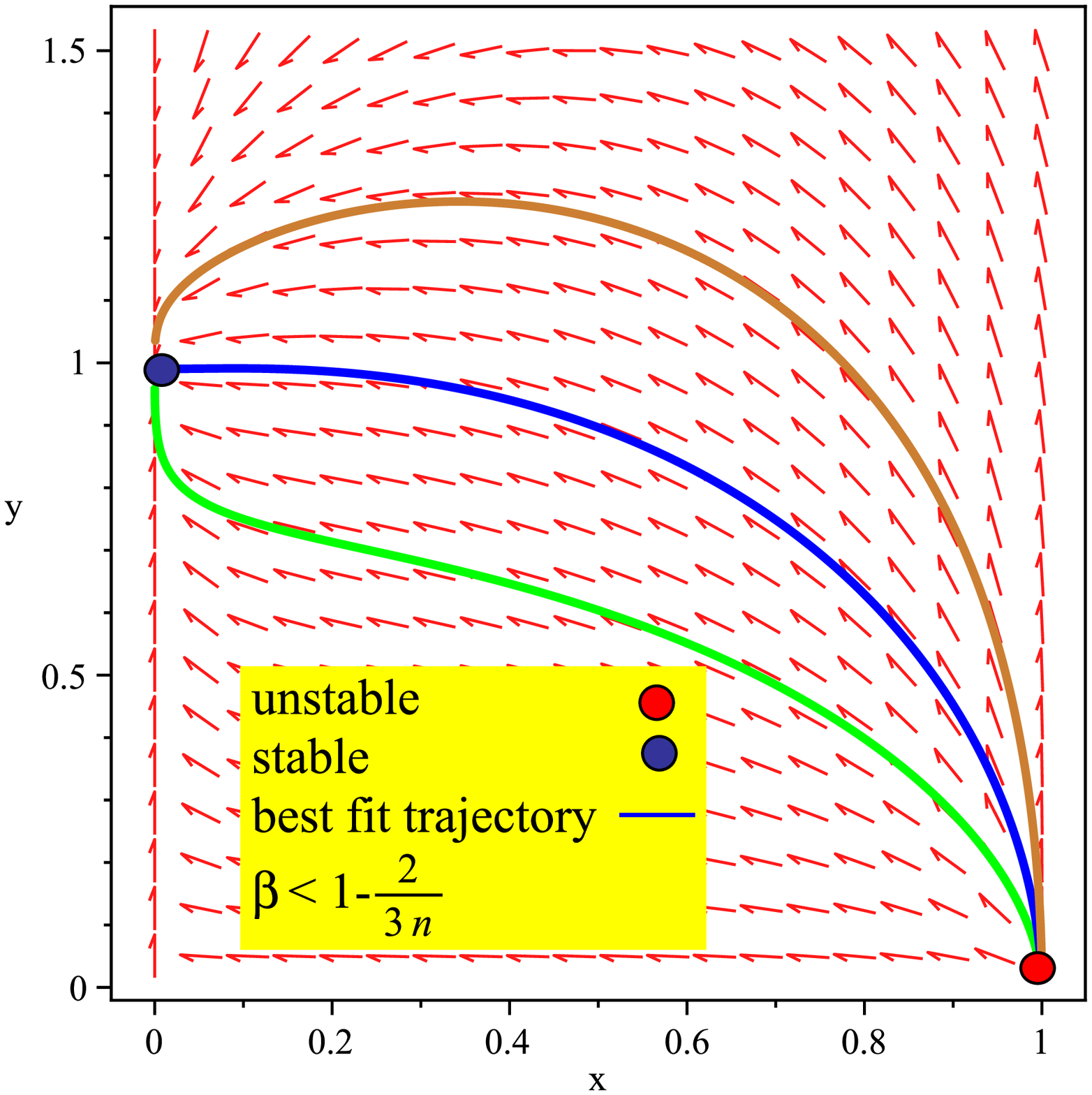}
\caption{Position of the fixed points of our model when $\beta>1-\frac{2}{3n}$ (left), and when $\beta<1-\frac{2}{3n}$ (right). The blue curve shows the best-fit trajectory.}\label{fig1}
\end{figure}

Eigenvalues have been expressed in terms of $\beta$ and $n$. It can be seen that point $A$ is always an unstable point irrespective of the value of $\beta$. But point $B$, is a saddle point if $\beta>1-\frac{2}{3n}$, and is a stable point if $\beta<1-\frac{2}{3n}$. Regarding TABLE \ref{table:1}, it is clear that for the best-fitted values of $\beta$ and $n$, point $B$, is a stable point. FIG.\ref{fig1}, demonstrates the phase portrait of our dynamical system in $xy$-plane for the best-fitted values.

\section{Big Rip Singularity}

One of the most important problems in cosmology is the so called big rip singularity. This problem was first brought up in DE models in which the DE component is a phantom fluid with $w_{DE}<-1$. In these models the energy density of the phantom fluid grows with the expansion of the universe so that it blows up at a finite time in the future and all bound structures rip apart \cite{Caldwell3} . Although some authors have shown that phantom big rip singularity can be prevented via interactions between the dark sectors of the universe \cite{Curbelo}-\cite{Fu} , this problem may still appear in some other modified gravity theories and also brane-world scenarios, because in these cases one can introduce an effective DE component with a dynamical EoS parameter $w_{eff}$, which may cross the phantom divide line ($w=-1$), and end up with a phantom phase in the future and as a consequence would yield the big rip singularity. So, in order for a cosmological model to be more viable, it should avoid this singularity.

The direct condition for the avoidance of big rip is $w_{tot}>-1$\cite{Wei2} . Although in an ADE model $w_{DE}$, is always greater than -1, when one considers an interacting ADE model, $w$-crossing occurs. But neither of the models suffers from the big rip problem, because $w_{tot}$ is always greater than -1. On the other hand, neither a $\Lambda$DGP model\cite{Chimento}, nor a SDGP\cite{Zhang2} model yield a big rip singularity, though in the latter $w$-crossing happens. In another article, the authors have shown that a DGP model with HDE, suffers from the big rip singularity\cite{Wu}. But the case is different for ADE in a DGP brane-world model. As it has been illustrated in \cite{Farajollahi4} , both the $w_{DE}$ and $w_{tot}$, are always greater than -1, and this model does not approach big rip singularity.

\begin{figure}
\centering
\includegraphics[width=8cm]{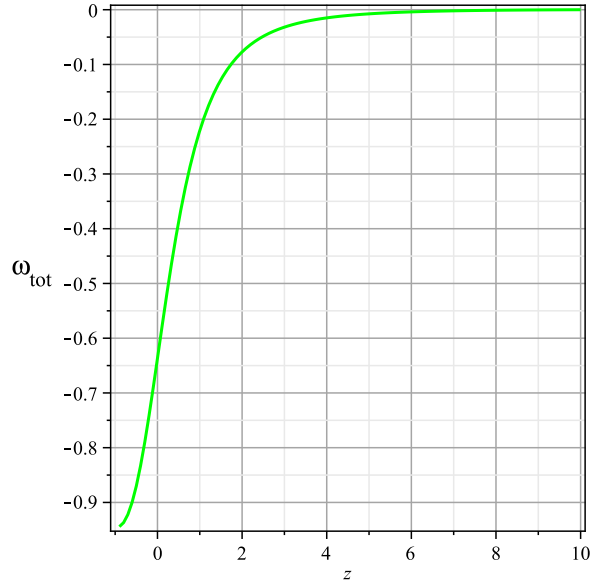}
\caption{The evolutionary curve of total EoS  parameter for the best-fitted values of combining CMB+BAO+OHD.}\label{fig3}
\end{figure}

To discuss the role of interaction in a DGP model in the presence of ADE, we write the total EoS parameter in terms of our new variables as
\begin{equation}\label{eostot}
    w_{tot}=-1-\frac{2\dot H}{3H^2}=-1+\frac{2z^2(x^2+\frac{2}{3n}y^3z-\beta x^2y^2)}{z^2+1}\,.
\end{equation}
With attention to Eq.(\ref{eostot}), we can estimate the value of $w_{tot}$, in different epochs of the universe in our model. It is easy to determine that in matter dominated era (point $A$) and also in DE dominated era (point $B$), $w_{tot}$, has a value greater than -1. What is important for us is the value of $w_{tot}$, in the future. This means that the future big rip singularity is avoidable in our model. Using the best-fitted model parameters, we have illustrated the behavior of total EoS parameter of the model. Figure \ref{fig3}, indicates that the universe undergoes acceleration without entering phantom regime in the past or future.

\section{Conclusion}

In this manuscript we investigated the effect of interaction between ADE and dark matter in a normal branch of DGP cosmology in the context of dynamical system approach. After introducing the model and the new phase-space variables, we obtained a two dimensional autonomous system. We first constrained model parameters numerically and using observational CMB+BAO+OHD data, and then tried to find critical points of the model. We found two critical points, an unstable point which was related to matter dominated era and also another DE dominated critical point which could be stable if $\beta<1-\frac{2}{3n}$. Hence, we concluded that interaction could lead to a stable universe in the future. This is very interesting, because as it has been mentioned in \cite{Arvin} , a non-interacting ADE model in DGP cosmology does not include any stable critical point. Thus, in addition to the known role of interaction in solving the coincidence problem we represented that it could be very useful in stabilization of the universe. Also, we indicated that there is not the big rip singularity in our model.

\end{document}